# Performance Analytical Comparison of Blockchain-as-a-Service (BaaS) Platforms

Md Mehedi Hassan Onik [1] and Mahdi H. Miraz [2,3]

[1] Department of Computer Engineering, Inje University, Gimhae 50834, Korea
hassan@oasis.inje.ac.kr
[2] The Chinses University of Hong Kong (CUHK), Sha Tin, Hong Kong
[4] Wrexham Glyndŵr University, Wrexham, UK
m.miraz@ieee.org

**Abstract.** Both blockchain technologies and cloud computing are contemporary emerging technologies. While the application of Blockchain technologies is being spread beyond cryptocurrency, cloud computing is also seeing a paradigm shift to meet the needs of the 4th industrial revolution (Industry 4.0). New technological advancement, especially by the fusion of these two, such as Blockchain-as-a-Service (BaaS), is considered to be able to significantly generate values to the enterprises. This article surveys the current status of BaaS in terms of technological development, applications, market potentials and so forth. An evaluative judgement, comparing amongst various BaaS platforms, has been presented, along with the trajectory of adoption, challenges and risk factors. Finally, the study suggests standardisation of available BaaS platforms.

## 1 Introduction

Blockchain, as first introduced in 2008 by Shatoshi Nakamoto [1], as the technology behind Bitcoin, has now matured enough as a standalone technology. Applications of blockchain have reached far beyond cryptocurrencies [2]. Examples of non-monetary applications of blockchain include securities settlement [3, 4], supply-chain [2], HR management [5], Healthcare [2, 6], decision making [7], personal data management [8, 9] and so forth [10]. In fact, blockchain is not a completely new technology, rather it just a new incorporated mechanism utilising multifaceted existing technologies together – such as distributed ledger technology (DLT), mathematical hashing, distributed networks, asymmetric encryption techniques, digital signatures and programming [11] – for the system to perform seamlessly. A transaction in a blockchain ecosystem is triggered by the sending node, verified and validated by the other participating nodes and if a consensus is reached it is then added to the pool of "unconfirmed" transactions to form a 'block'. The creation of the block varies depending on the consensus algorithm (e.g. Proof-of-Work, Proof-of-Stake etc.) used. However, once a block is successfully formed, is then propagated to all the nodes in the network to be appended at the open end of their existing copy of the chain. Thus, all the distributed copies of the database (ledger) is updated and synchronised. Because blocks are mathematically bound by cumulative hashes, altering a single transaction or even a single bit will invalidate that block and rebuilding block with a new hash will invalidate the following blocks. Thus, it acts as a "Trust Machine" [12] which brings immutability, security, eliminates single point of failure (SPF) as well as the need for a third-party for establishing trust.

Cloud computing has been defined differently by different bodies or professionals. However, the definition of cloud computing provided by the national institute of standards and technology (NIST), part of the U.S. Department of commerce—in its Special Publication 800-145 [13], is the widely accepted one. NIST [14] defines cloud computing as "*a model for enabling convenient, on-demand network access to a shared pool of configurable computing resources (e.g., networks, servers, storage, applications, and services) that can be rapidly provisioned and released with minimal management effort or service provider interaction.*" Further to this definition, cloud computing enables a model of IT service in any combination of IT resources – from a network accessible data storage to a fully-fledged virtual machines, from hosted application/service to application/service development infrastructure [15, 16]. A cloud consumer can simply avail the required resources from the pool through service orchestration. The resources are released and returned to the pool when the consumer no longer needs them. The cloud model functions analogous to regular utility services such as electricity. When required, a consumer plugs in the appliance into a socket and switch it on – in most cases without knowing the details of how electricity is produced and distributed. The consumers are only charged for whatever amount of electricity they have consumed. In a similar way, the cloud model abstracts the IT infrastructure for enabling the consumers to rent IT resources eliminating the associated costs and risks of owning these resources. However, the cloud is not limited to infrastructure, rather offers platforms, services and applications too making cloud service even more pervasive. Finally, cloud converts capital expenditure (CapEx) and operational expenditure (OpEx) making it popular among the small to medium enterprises.

One of the next generation cloud computing features is Blockchain-as-a-Service (BaaS) – a fusion of blockchain technology and the cloud computing model. BaaS enables offshoring the implementation of blockchain for any enterprise to the cloud environment, without needing any IT expertise. Thus, enterprises can benefit from BaaS as a utility service and serve their business need. BaaS relatively being a new addition to both blockchain and cloud technologies, this article conducts an extensive survey of relevant research literatures and projects as well as performs a performance analytical comparison of Blockchain-as-a-Service (BaaS) platforms- Mainly those provided by Microsoft, Amazon, Hewlett Packard (HP), Oracle and SAP. The paper also discusses future challenges, risk factors and trajectory of adoption.

## 2 Overview of Blockchain-as-a-Service (BaaS)

### 2.1 Overview

**Blockchain:** The blockchain technologies utilise decentralised distributed ledgers for recording the transactions across a peer to peer network. Without being dominated by any central authority and/or middle man for "trust", this technology can verify, validate and complete transactions being autonomously governed by the coded protocol and consensus approach powered by the nodes of the peer-to-peer networks. In fact, blockchain technology was first introduced in 2008 as a core technology for a cryptocurrency (i.e. Bitcoin) [1]. However, successful applications of BC in multifaceted other usecases beyond cryptocurrencies have instituted it as one of the cardinal technologies of both the emerging and the upcoming industrial revolutions [2–10] evident by the forecasted business value creation of blockchain technology to exceed \$3.1 trillion by 2030 [17]. Based on the level of write and read access to the ledger, blockchain can be categorised into three types: public (permissionless) blockchain – mainly for cryptocurrencies (e.g. Bitcoin, Ethereum), private (permissioned) blockchain – mainly for non-monetary applications within a closed network and Federated (consortium/hybrid) blockchain – a combination of both public and private mainly to be used within a consortium (e.g. Hyperledger). Anyone at any time can join and leave the public blockchain ecosystem enjoying full access to read and write (subject to consensus). Joining in a private blockchain is restricted - read and write access are controlled based on the roles of the nodes or other restrictions as imposed by the protocol. In a hybrid blockchain, joining is sometimes controlled by invitations only – while all the participating nodes enjoy read access as in public blockchain, write access is limited as in private blockchain.

**Blockchain-as-a-Service:** Blockchain-as-a-Service (BaaS) means of building, managing, hosting and using various aspects of blockchain technologies such as applications, nodes, smart contracts and distributed ledger, on the cloud. Such cloud-based service facilitates blockchain set-up, platform, security and other associated features. Thus BaaS introduces the blockchain service platform, supporting blockchain core features, based on cloud computing infrastructure with the integrated developing environment for both the developers and the consumers [18–20].

In fact, the key concept of BaaS is almost similar to that of Software-as-a-Service (SaaS). According to cloud computing orchestration, BaaS can function either explicitly utilising Platform-as-a-Service (PaaS) or implicitly via Software-as-a-service (SaaS). Based on how it is implemented, the locus of BaaS in a cloud computing environment may vary. Fig.1 demonstrates the location of BaaS in an on-premise local implementation. In such implementations, BaaS functions with the support of both SaaS and PaaS. While BaaS receives the technical services (software) from SaaS, it gets infrastructure support from PaaS. On-premise blockchain implementation is highly expensive. Users of such local implementation require investing a significant share of capital expenditure (CapEx) for maintaining the infrastructure and performance of the DLT. The alternative economical approach is BaaS – a user can enjoy full service of the blockchain technology even investing less. BaaS can manage blockchain consensus, forking, node validity, commodity exchange, backup, off-chain and on-chain synchronisation all by itself. Similarly, BaaS can also manage resource, bandwidth, internet connection and other associated services. However, BaaS provides the enterprises with the flexibility to emphasise on business logic and functional need of the blockchain. BaaS helps to create, develop, test, host, deploy and operate blockchain related applications on cloud infrastructure. BaaS implementation fully out-sources the technical overhead to the cloud service provider. Fig. 2 shows the architectural overview of BaaS.

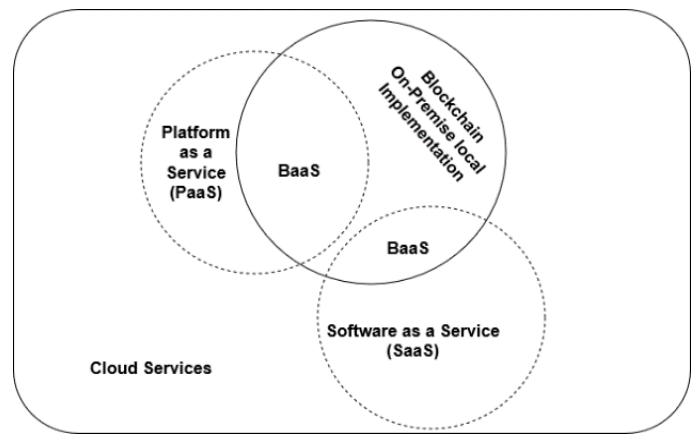

**Fig. 1.** Location of BaaS in compare with other cloud services

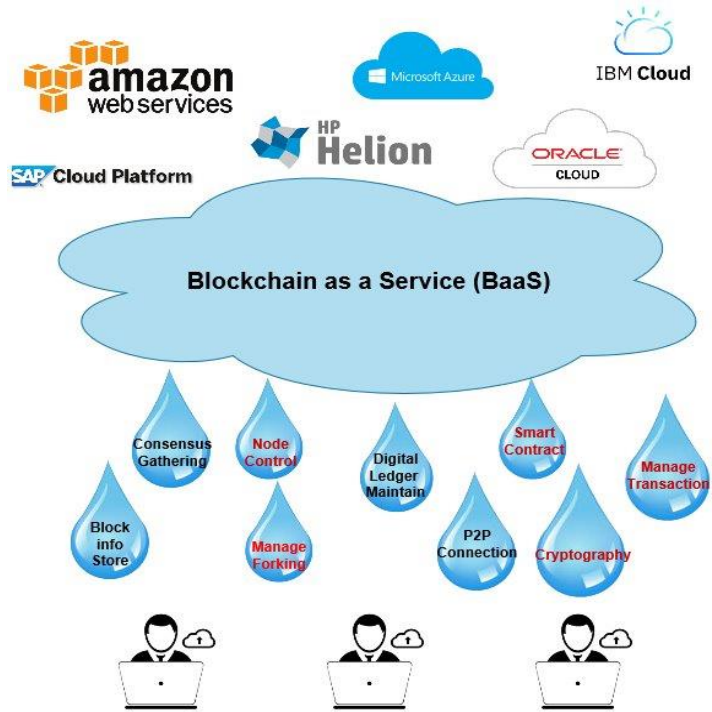

**Fig. 2.** Architectural Overview of Blockchain-as-a-Service (BaaS)

## 2.2    Advantages of BaaS:

Real-world blockchain use cases are rapidly emerging, but the skills and resources required to build blockchain applications are neither widely available nor cheap. Therefore, BaaS possesses the potentials to address this aperture and make blockchain technology accessible to a broader audience. A few benefits of blockchain-as-a-service (BaaS) are discussed here:

- With already established cloud platform blockchain adopters can receive seamless service with far more fewer cost than actual (on-premise) implementation.
- In current blockchain architecture, several regulations and norms like node verification, node attachment, node deletion, forking must be taken care off. However, BaaS can take care of them without any intervention.
- Blockchain technology is being used beyond cryptocurrencies. Therefore, interaction with another platform, service, infrastructure has increased a lot in the last few years. Since BaaS blockchain technology is built utilising existing cloud infrastructure, PaaS, IaaS, SaaS and similar other aspects of the cloud remains native to BaaS – offering a higher degree of interoperability.
- Current blockchain implementation requires a moderate degree of knowledge in the domain of cryptography and distributed technologies. Alternatively, BaaS, which is offered as a complete service by the providers, allows deploying, managing and operating of enterprise blockchain technology without any technical knowledge.

## 3 Overview of Available BaaS Platforms

### 3.1 Microsoft Azure BaaS:

During late 2015, Microsoft aligned with Consensys to offer Ethereum Blockchain-as-a-Service (EBaaS) [21]. As Microsoft corporation already possessed a widely used infrastructure and cloud platform (i.e. Azure), coupling up blockchain technology as a service on their existing Azure platform was a rational business move. In order to offer BaaS, 'Azure Blockchain Workbench' was introduced with two major tools: 'Microsoft Flow (Ether.Camp)' and 'Logic Apps (BlockApps)'. The aforementioned establish a scalable and integrated blockchain development environment along with a consortium Ethereum blockchain application development environment. Azure Blockchain Workbench (ABW) allows direct development of distributed applications (DApps) without worrying much about the underlying system services. With the available REST APIs, ABW facilitates the users to integrate other available services to interact with the newly created personalised application. ABW has the ability to connect available Microsoft services like office 365, Excel, SharePoint, 365 CRM and other available services. More than 200 connectors are considered to provide a graphical user interface in 'Logic Apps' and 'Flow' which minimise end to end blockchain management complexity [20, 22, 23].

The ABW fully complement legacy Blockchain technology and provide core blockchain services. Identity management is ensured with the help of the Azure active directory. The Azure Blockchain Workbench also manage both the user roles and the smart contract. It allows the users to write their own access and business logic code (smart contract). Finally, for privacy-preserved data mining, ABW synchronised on-chain information with off-chain SQL server (on demand). This empowers the data analysing the scope of ABW many times. In addition, for seamless interaction amongst available software services, Microsoft Azure also provides Azure Blockchain Development Kit (ABDK). Microsoft ABDK offers linking interfaces, assimilating data and systems, deploying blockchain networks. ABDK can interact with legacy applications and protocols like FTP, Microsoft Excel, email data. Several legacy databases such as SQL, Excel, PowerBi and Azure Search service as well as other SaaS deployment such as SharePoint, Dynamics and office 365 can also be accessible by ABW though ABDK. Key advantages of ABW are as follows [20, 22, 23]:

- Using ABW, configuration, deployment and testing of any BaaS application in a consortium network can be performed by only a few clicks. ABW's by default ledger deployment and network infrastructure reduces infrastructure creation period.
- Overall blockchain technology development time and the cost are reduced by making proper use of Azure cloud services such as Azure Active Directory (AD) for easier sign-in and identity checking, storing private keys with Azure Key Vault, secure and easy messaging among blockchain nodes, off-chain and on-chain data synchronisation for privacy-preservation and visualisation.

- ABW facilitates easy integration between any business entity and the blockchain technology. With Microsoft's ABW and ABDK (REST-based API) interaction, messaging, verifying with blockchain nodes (clients) have become much easier than before.
- Finally, Microsoft acquires comparatively more platforms, services and infrastructures than any other cloud providers. This leaves a company with higher success and lower compatibility issues.

### 3.2 Amazon AWS BaaS:

Initially, Amazon started providing blockchain by partnering with third parties (R3, Kaleido) [24, 25], however, it recently announced its own blockchain platform. Later, Amazon declares its own blockchain service based on Hyperledger in two different forms: Amazon Quantum Ledger Database (QLDB) and Amazon managed Blockchain. In addition, Amazon's AWS provides developers with a wide selection of blockchain frameworks with minimum pricing [24, 25].

**Blockchain Amazon Quantum Ledger Database (QLDB):** Amazon QLDB is a new database that provides the functionalities of a distributed ledger database without creating a ledger. Amazon QLDB mainly focused on developing an immutable and transparent ledger. This QLDB can create a distributed ledger application both with relational and blockchain database. To maintain both immutable (relational) and distributed (Hyperledger Fabric and Ethereum) databases simultaneously, individual blockchain node along with the network must be validated. In order to track every data exchange among blockchain nodes, QLDB maintains a ledger named Journal. It's an immutable transaction log where transactions are saved as a new block. In addition, the journal determines current and history of all the transactions [24, 25]. Fig. 3 demonstrates the architecture of Amazon Quantum Ledger Database for BaaS.

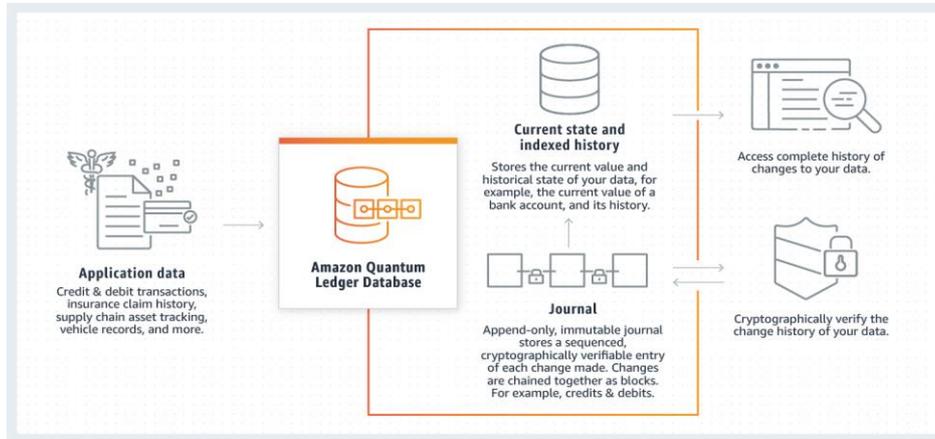

**Fig. 3.** Amazon QLDB (BaaS) [24–26]

**Amazon Managed Blockchain:** Amazon Managed Blockchain (AMB) is a blockchain network backed by the Hyperledger fabric. A full network can be installed within 10-15 minutes. It's a private network meant solely for blockchain based technologies. However, most of its functionalities are the same as QLDB [24–26]. Fig. 4 presents the basic architecture of Amazon Managed Blockchain as a BaaS.

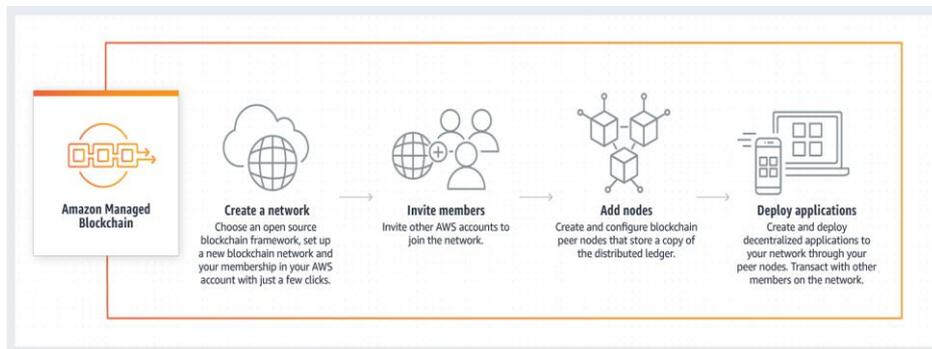

**Fig. 4.** Amazon Managed Blockchain (BaaS) [24–26]

### 3.3    IBM BaaS:

IBM revealed BaaS in the year of 2017 using the Hyperledger fabric on IBM cloud. This allows any private and public organizations to introduce private, public or consortium blockchain. IBM also introduced a 'SecureKey Technologies', a digital identity sharing key to protect the public-private key. IBM claims its 'blockchain-as-a-service' technology to be highly auditable and performs better than other SaaS services [27–29].

IBM provides SaaS through 'Bluemix' [27–29]. With the help of the 'Bluemix', developers are allowed to create blockchain application without any extra setup. With the aid of 'Bluemix', 'Hyperledger Fabric' and IBM cloud users can directly develop a DevOps and deploy Chaincode. Chaincode is a software used by IBM to maintain business logic (consensus) and can be written with Go and Node.js. IBM blockchain has 'Transactor's who are actually acting as clients using application programming interfaces (API) and software development kit (SDK). IBM also introduces the concept of the validating peer (VP) and non-validating peer (NVP). Only VP are able to participate in IBM SaaS directly. Alternatively, NVP can also be connected with the chain via REST API. However, for security reason, NVP can only forward a request to a VP rather than performing the actual work. High-level architecture of IoT applications that use IBM Cloud-based Hyperledger services is shown in Fig. 5. As of now, IBM has two versions of BaaS (1.0 and 2.0). IBM BaaS 2.0 [30] is comparatively more robust and offer the following benefits:

- The current version of IBM BaaS (2.0) allows large scale development, extensive test and public production in a single BaaS environment.
- The IBM Blockchain platform supports the smart contracts to be written in three popular languages such as JavaScript, Go and Java.
- Operation, governance and deployment of the blockchain components are solely controlled by the users.
- IBM BaaS nodes can operate in any environment such as private, public and hybrid clouds.

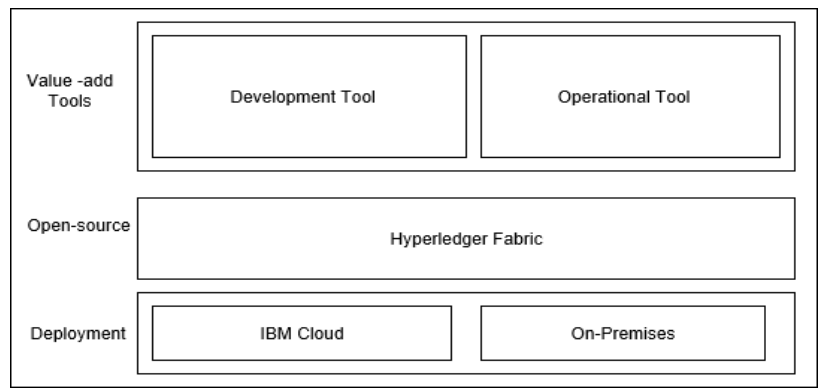

**Fig. 5.** IBM BaaS [30]

### 3.4 Hewlett Packard (HP) BaaS

Hewlett Packard Enterprise (HPE) introduced their first ever BaaS named 'Mission Critical Distributed Ledger Technology' (MCDLT) or DLT as a Service [31]. HPE's MCDLT includes higher scalability and SQL integration with blockchain technology. This solution includes replacing on-premise user infrastructure with public cloud envi-ronment or generic infrastructure. HPE partnered with R3 (software company) to establish a 100% fault tolerance blockchain application development platform for enterprise use.

### 3.5 Oracle BaaS:

Oracle recently introduced Oracle Blockchain Cloud Service (OBCS) besides their already established Platform as a service (PaaS) and Software as a Service (SaaS) [32]. In order to start the internal blockchain (distributed ledger) project quickly, Oracle BaaS introduced two key concepts [33]. Firstly, OBCS possesses turn-key sandbox which is solely designed for the developers. Secondly, independent software vendors (ISV) facilitates easy deployment of blockchain technology regardless of their vendor (Fig. 6).

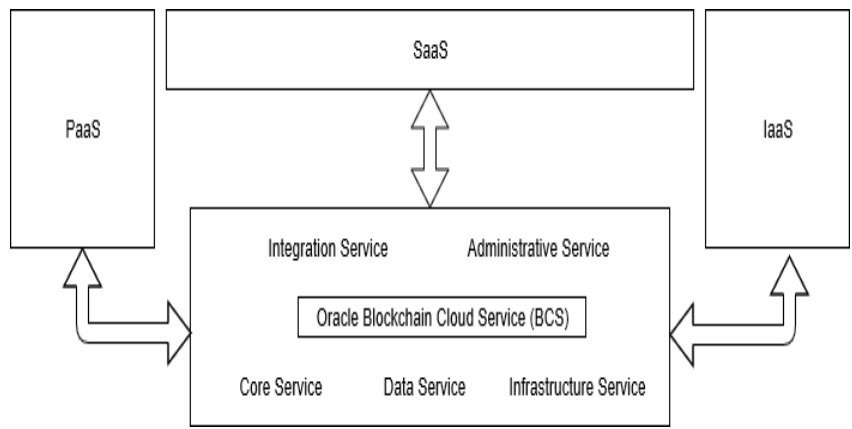

**Fig. 6.** Oracle BaaS [33]

### 3.6 SAP BaaS:

A great addition to BaaS is SAP Blockchain [34]. SAP introduced both SAP-Cloud-Platform Blockchain Service and SAP HANA Blockchain Service. SAP HANA connects any SAP HANA database to the most popular enterprise blockchain platforms [35]. This provides very interesting capabilities that were previously unheard of in the blockchain ecosystem.

SAP HANA blockchain (BaaS) connects the SAP HANA database with distributed ledger technology (DLT). SAP HANA supports stellar consensus protocol (SCP) blockchain within it. SCP blockchain can be hosted on any third-party cloud and local infrastructure. In addition, SAP HANA cloud services are only available if it is hosted on SCP. SAP HANA is not a blockchain node, rather it configures the connection properties of SCP. SAP HANA BaaS maintain the blockchain transaction details in 3 types of SAP HANA database tables, as shown in Fig. 7:

- Blocks and transactions information is saved as 'Raw data'.
- History of transactions along with the messages is kept in a ledger.

- Latest valid tuples of a blockchain transaction are saved in 'Worldstate'.

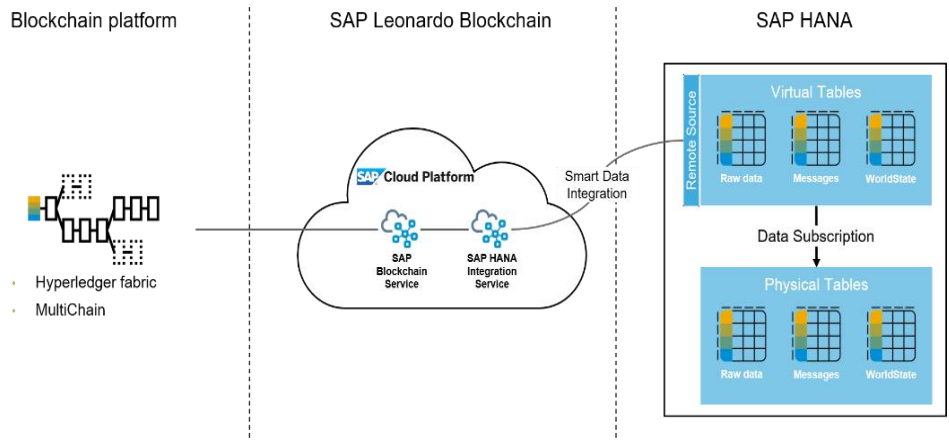

**Fig. 7.** SAP HANA BaaS [34]

## 4 Comparison of BaaS Platforms

From the aforementioned sections with few other studies [36–41], we compare the performance of the available BaaS platforms. Table 1 provides availability of the several blockchain hosting platforms by top BaaS platforms:

**Table 1.** BaaS platform vs Hosting platform availability comparison

|  | Ethereum | Quorum | Corda | Hyperledger Fabric | Multi-chain | Digi-tal As-set |
|---|---|---|---|---|---|---|
| AWS | √ | √ | √ | √ |  |  |
| Azure | √ | √ | √ | √ | √ |  |
| Google | √ |  |  | √ |  | √ |
| HPE |  |  | √ |  |  |  |
| IBM |  |  |  | √ |  |  |
| Oracle |  |  |  | √ |  |  |
| SAP |  |  |  | √ |  |  |

In Table 2, other aspects are compared to available BaaS services. Available partners, major users, authentication mechanism, pricing, blockchain access type, development facility and scalability factors are compared amongst key BaaS platforms.

**Table 2.** Comparison of BaaS platforms

| | Azure | AWS | IBM | Oracle | SAP |
|---|---|---|---|---|---|
| Major Partner | Corda, Blockapps, GoChain, Consensys | Cisco, Intel, Keleido, Corda, R3, Blockapps | SecureKey Technologies, Canadian banks | Tron, Aurora, Steemit, Pantera | Intel, UPS, HPE, Airbus |
| Major User | Xbox, 3M and Insurwave | T mobile and Guidewire | Arab Jordan Investment Bank, CargoSmart, Certified Origins, Intelipost, Nigeria Customs. | commercial bank Banco de Chile, Circulor, SERES, and CDEL, HealthSync | |
| Authentication and authorization | Active Directory | Identity and Access Management | IBM Secured Services Containers | Identity federation | Service Key |
| Pricing | Subscription plan and pay as per sue | Pay as per use | Monthly Subscription, Free trial | $0.75 pay as you go | |
| Blockchain type | Permissioned | Permissioned | Permissioned | Permissioned, Consortium | Permissionless |
| Development Facility | High with Microsoft development kit | Medium, limited only with AWS kit | IBM Bluemix development platform | Hyperledger Fabrik SDK | Not yet Released |
| Scalability | High with all Microsoft products | Provide API for quick node creation | IBM Smart Cloud only | | |

In summary, as Azure and AWS have already established cloud infrastructure, they are in a strong position than other services. Alternative, an increase of on-premise (local database) use intensify the usage of Oracle, IBM and SAP services. However, service provided by Azure and AWS is costly while SAP provides relatively cheaper service. As BaaS platforms' security, cost and efficiency are changing rapidly, a stable release of enterprise BaaS platforms will open more scopes for comparison. Next section will discuss the future scopes, research directions and recommendations to help choosing efficient BaaS platform.

## 5 Future Research Challenges and Risk Factors:

Three major problems of blockchain technologies, as inherited in its architecture, are lack of scalability [3, 42], lack of interoperability [3, 43] and its antithetic stand against the notion of green computing [44, 45]. On the contrary, despite its widespread adoption, cloud computing also suffers from varies limitations such as lack of standardisation leading to vendor lock-in, security and privacy concern as well as data ownership and locality issue. While both the technologies are relatively immature, integration of both may give birth to new complexities in terms of technical aspects.

Blockchain highly suffers from scalability problem due to its capped transaction latency as well as consensus approach – as injected in its architecture to provide better security and to eliminate double spending problem [42]. Many research have been conducted so far to overcome this issue, keeping the base technology unaltered as it has already been proven to be highly secure. Recent advancement in the development of Bitcoin's lightning networks (LN) and similar technologies forecasted to play a vital role in addressing this issue. In a LN [42], direct transactions between two parties can take place in a tête-à-tête fashion, via a payment channel constructed in a separate (second) layer on top of the base layer of the chain. These transactions are considered as intermediate transactions which are not subject to normal consensus approach. As a result, the transactions are "instantaneous". That being said, they are still relatively slow compared to fiat currency transactions such as those facilitated by Visa. While the intermediate transactions broadcasted to the nodes of the peer-to-peer network for consensus, the final balance needs to be validated and verified by the nodes for settlement on the base chain once a channel is closed. With the help onion style routing, it is possible to perform LN transactions amongst the peers who are not "directly" connected by any LN channel between themselves, while maintaining the same level of privacy. However, the success of LN depends on the level of future technological maturity as well as the rate of adoption. BaaS can play an important role in this regard, by implicitly increasing the LN adoption trend.

Because tokens or coins exist only on their respective native chains, there is no straightforward method to swap two different tokens or coins or (transaction) data of different blockchain ecosystems. However, the recent development of Atomic Swaps [43] holds the potentials of addressing the interoperability problem to some extent. The term "Atomic" has been taken from database systems where atomicity means an operation (i.e. swap of two different cryptocurrencies in this case) will happen either completely or not at all. LN network powered atomic swaps also support off-chain scaling. Thus, both LN and Atomic Swaps together – if the technologies mature as expected – possess great potentials to accelerate BaaS adoption.

Most of the blockchain consensus approaches, including the most widely use Proof-of-Work (PoW), demand a tremendous amount of power consumption. Thus, its antithetic stand against the notion of green computing is highly critiqued [44]. However, if the tasks associated to consensus are outsourced from the cloud nodes via BaaS, which are already being run anyway, can save the "extra" demand of electricity needed for this purpose [45].

One of the major problems of cloud computing, as stated above, is the lack of standardisation leading to vendor lock-in i.e. lack of interoperability and portability. A vendor lock-in takes place when altering the cloud service provider becomes either impossible or highly expensive. Such situations mostly happen when there are non-standard proprietary services offered by the cloud service providers or if there are no viable alternatives. With the maturity of the cloud technology, while "generic" cloud services today is far more standardised than it was in the past, this is not the case for specific cloud services such as BaaS. Thus, lack of standardisation may still remain as a major challenge and risk factor for BaaS. While technologies like atomic swaps may be applied to address this problem, the viability has not yet been measured and it remains uncertain concerning to what extent atomic swaps can help.

Since BaaS is not primarily aimed to facilitate cryptocurrency transactions, rather the targeted applications are in the domain of non-monetary data transactions and storage, it is not going to be a subject to money-laundering or other financial regulations. However, both blockchain and cloud computing being distributed in nature, they are subject to data ownership, data localisation and data privacy regulations, especially in regards to EU General Data Protection Regulation (GDPR) and similar other regulations in various legal jurisdictions [42, 43].

To surmise, cloud computing, blockchain technologies and the fusion of both i.e. Blockchain-as-a-Service (BaaS) are still considered as immature. Thus, the fusion possesses significant risk factors and the future adoption trend of it significantly depends on many aspects including legal and regulatory ones.

Finally, the study suggests the following key considerations while choosing a BaaS platform:

1. Feasibility of the BaaS platform to solve real-world problems.
2. Scalability of the BaaS platform to host ever-increasing hosts (nodes).
3. Availability of the community support of a BaaS platform.
4. Feasibility of the BaaS platform from coding or modification perspectives.
5. Adaptability with the existing technologies.
6. Accessibility (public, private or consortium) of a BaaS platform.
7. Security and privacy of a BaaS platform.

# 6    Conclusions:

By briefly introducing both the blockchain and the cloud computing technologies, the paper then presents the concept of Blockchain-as-a-Service (BaaS) – the fusion of both the technologies. A comprehensive survey of the current status of BaaS in terms of technological development, applications, market potentials and so forth was also presented. To form an evaluative judgement, the paper also compared various BaaS platforms such as Microsoft Azure Ethereum Blockchain-as-a-Service (EBaaS), Azure Blockchain Workbench - Microsoft Flow (Ether.Camp) and Logic Apps (BlockApps), Amazon AWS, Amazon Quantum Ledger Database (QLDB), Amazon Managed Blockchain, IBM BaaS, Hewlett Packard (HP) Mission Critical Distributed Ledger Technology (MCDLT), Oracle Blockchain Cloud Service (OBCS), SAP-Cloud-Platform Blockchain Service and SAP HANA Blockchain Service. The paper also attempts to forecast the trajectory of adoption of BaaS and its challenges as well as risk factors. Finally, future research directions are outlined.

In future, our goal is to establish an access control aware personal information access platform with BaaS architecture. In addition, future studies will consider R3, HPE R3, BitSE, Blocko, PayStand, Blockstream and other BaaS platforms. In our future studies, energy efficiency and privacy preservation in blockchain technology will be our main concern.